**Parent grain reconstruction from partially or fully transformed microstructures in *MTEX***

Frank Niessen[1,*], Tuomo Nyyssönen[2], Azdiar A. Gazder[3], Ralf Hielscher[4]

[1]*Technical University of Denmark, Department of Mechanical Engineering, 2800 Kgs. Lyngby, Denmark*

[2]*Swerim AB, 164 40 Kista, Sweden*

[3]*Electron Microscopy Centre, University of Wollongong, New South Wales 2500, Australia*

[4]*Technische Universität Chemnitz, Fakultät für Mathematik, 09126 Chemnitz, Germany*

# Abstract

A versatile generic framework for parent grain reconstruction from fully or partially transformed child microstructures was integrated into the open-source crystallographic toolbox *MTEX*. The framework extends traditional parent grain reconstruction, phase transformation and variant analysis to all parent-child crystal symmetry combinations. The inherent versatility of the universally applicable parent grain reconstruction methods, and the ability to conduct in-depth variant analysis are showcased via example workflows that can be programmatically modified by users to suit their specific applications. This is highlighted by three applications namely, α′-to-γ reconstruction in a lath martensitic steel, α-to-β reconstruction in a Ti alloy, and a two-step reconstruction from α′-to-ε-to-γ in a twinning and transformation -induced plasticity steel. Advanced orientation relationship discovery and analysis options, including variant analysis, is demonstrated via the add-on function library, *ORTools*.

[*]Corresponding author, Contact: frannie@mek.dtu.dk

# 1 Introduction

Depending on its chemistry and thermo-mechanical processing history, the free energy difference of a material system may lead to the phase transformation of a pre-existing parent phase with a given crystal symmetry to a new child phase with a similar or different crystal symmetry [1]. In many instances, this phase transformation occurs via the operation of an orientation relationship (OR). An OR refers to the coherent geometric parallelism between specific planes and directions of parent-child crystal symmetries on either side of their common boundary segment [2]. In all parent-child crystal symmetry combinations, one or more favored ORs exist that provide the best fit at their boundary segment interfaces and enables crystallographic phase transformation between them.

In the late 1800s, Adolf Martens discovered the now well-known martensitic transformation in steel. It involves a phase transformation from face-centered cubic (fcc) parent austenite ($\gamma$) to body-centered tetragonal (bct) child martensite ($\alpha'$) via the operation of six or more ORs. Of these, the Bain, Kurdjumov–Sachs (K-S) [3], Nishiyama–Wassermann [4], Greninger–Troiano, Headley–Brooks and Pitsch ORs are the most frequently reported. To serve as an example, the K-S OR between austenite and martensite is expressed as: $\{111\}_\gamma || \{1\bar{1}0\}_{\alpha'}$, $\langle\bar{1}10\rangle_\gamma || \langle\bar{1}11\rangle_{\alpha'}$. In this case, the $\langle\bar{1}10\rangle_\gamma$ and $\langle\bar{1}11\rangle_{\alpha'}$ directions lie in the $\{111\}_\gamma$ and $\{1\bar{1}0\}_{\alpha'}$ planes, respectively. The crystal symmetries of austenite and martensite are such that when the K-S OR is operative, a parent $\gamma$ grain with a given orientation may transform to any of 24 uniquely oriented child $\alpha'$ grains. The latter are referred to as child orientation variants, or more simply, as variants. Depending on the mechanism of phase transformation, variant selection may occur when only a few of the theoretically predicted child orientation variants dominate the partially or fully transformed microstructure.

The martensitic phase transformation has been subsequently harnessed to increase the mechanical strength of a wide range of commercial alloys [5,6]. The increase in mechanical strength is obtained by a Hall-Petch type hardening effect where the hierarchical subdivision of parent grains into child martensitic domains results in a decrease in the effective grain size of the alloy [7,8]. Depending on the alloy type, further strengthening may be achieved by interstitial solid solution strengthening [9,10] and/or increasing the dislocation density in the retained parent and child phases by plastic strain to accommodate the volume mismatch between them [11].

The above example of martensitic transformation is representative of a diffusionless displacive mechanism and signifies that: (i) the chemical composition of parent and child phases are similar, (ii) the atoms in parent and child unit cells maintain their sequence and atomic correspondence, and (iii) a shape change consistent with the parent-child crystal symmetry combination requires accommodation [2]. On the other hand, phase transformations that operate by diffusion and growth -based mechanisms occurring at relatively higher temperatures may also involve an OR. Such mechanisms involve elemental diffusion based on their preferential solubility in parent or child phases, which in turn, leads to differences in chemical composition and the loss of atomic correspondence between them. An often-cited example of such a mechanism is the parent β -to- child α phase transformation during the cooling of Ti and Zr alloys. Regardless of their displacive or diffusional origins, phase transformation mechanisms tend to involve ORs and variants.

For most of the twentieth century, uncovering ORs and analyzing local groupings of child orientation variants was only possible by laborious manual X-ray diffraction [3,4] and transmission electron microscopy [12] work. The continuous development of electron backscattering diffraction (EBSD) [13] from the 1980s onwards has enabled the mapping of entire microstructures and the characterization of a statistically significant number of parent and child orientations and morphologies [14].

With the widespread adoption of EBSD as a routine materials characterization technique, the first algorithms to reconstruct the parent phase from child orientation variants were concurrently developed to enable: (i) the analysis of parent phase microstructures prior to transformation, and (ii) variant analysis in the case of fully transformed microstructures. In 1994, Humbert et al. focused on child α -to- parent β reconstruction in Ti and Zr alloys [15] and in 2006, Cayron et al. [16] was the first to reconstruct parent austenite from child martensite in steel. Since then, various reconstruction algorithms were developed (the examples include but are not limited to Refs. [17–26]), with each new or modified iteration claiming superior performance in reconstruction accuracy and/or computational efficiency compared to previous implementations.

Of the many strategies for parent grain reconstruction that have been presented in the scientific literature to-date, several have tended to be proprietary software solutions for a limited number of ORs and/or parent-child crystal symmetry combinations. Consequently, the motivating factors for the present work were: (i) the implementation of a generic framework for parent grain reconstruction in the open-source crystallographic toolbox *MTEX* [27] and, (ii) the extension of parent grain reconstruction, phase transformation and variant analysis to all parent-child crystal symmetry combinations. The implementation features different parent reconstruction methods that may be called on and combined by users to create individual workflows to obtain a confident reconstruction of parent phase microstructures.

Following a brief overview of the common approaches to parent grain reconstruction, this work introduces the new parent grain reconstruction features in *MTEX* and the add-on software suite *ORTools* [28]. The latter comprises tools for OR discovery and analysis as well as variant analysis, and plots publication-ready figures of microstructures undergone partial or full transformation.

The parent grain reconstruction and analytical capabilities are demonstrated for three example alloys namely, (i) α′-to- γ reconstruction in a lath martensitic steel [20], (ii) α -to- β reconstruction in a Ti alloy, and (iii) a two-step reconstruction from α′-to- hexagonal close-packed (hcp) ε-martensite -to- γ in a twinning and transformation -induced plasticity steel [33].

## 2 Theory of parent grain reconstruction

In the following section, a parent orientation, $_S\mathbf{R}_P$, is a rotation matrix, **R**, that describes the coordinate transformation from the parent crystal basis, P, into the specimen coordinate basis, S. The formation of a child orientation, $_S\mathbf{R}_C$, via phase transformation of the parent orientation is then characterized by a misorientation, $_P\mathbf{R}_C$, that transforms the child coordinates into parent coordinates:

$$_S\mathbf{R}_C = {_S\mathbf{R}_P}\,{_P\mathbf{R}_C} \qquad (1)$$

As an example, a body-centered cubic (bcc) parent β grain with a cube orientation in a Ti or Zr alloy is defined by the following syntax in *MTEX*:

```
1 csP = crystalSymmetry('432',[3.3 3.3 3.3],'mineral','Beta');
2 oriP = orientation.cube(csP)
```
Code 1

The Burgers OR, $_C\mathbf{R}_P$, towards the hcp child phase α is defined as:

```
1 csC = crystalSymmetry('622',[3 3 4.7],'mineral','Alpha');
2 CRP = orientation.Burgers(csP,csC)
```
Code 2

With $_P\mathbf{R}_C = {_C\mathbf{R}_P^{-1}}$, the resulting child orientation is computed as:

```
1 oriC = oriP*inv(CRP)
```
Code 3

If $k$ = 1,..,K symmetry operators, $\mathbf{S}_P^k$, for the parent phase and $\ell$ = 1,..,L symmetry operators, $\mathbf{S}_C^\ell$, for the child phase are applied on the OR, it results in K × L symmetrically equivalent ORs, $\mathbf{S}_C^\ell\,{_C\mathbf{R}_P}\,\mathbf{S}_P^k$. For a particular parent orientation, $_S\mathbf{R}_P$, these produce a maximum K number of symmetrically non-equivalent child orientation variants:

$$_S\mathbf{R}_C^k = {_S\mathbf{R}_P}\,\mathbf{S}_P^k\,{_P\mathbf{R}_C} \qquad (2)$$

In cases when the OR transforms certain parent symmetries, $\mathbf{S}_P^{\hbar}$, into child symmetries, $\mathbf{S}_C^{\ell} = {}_C\mathbf{R}_P \mathbf{S}_P^{\hbar} {}_P\mathbf{R}_C$, a degenerated number of child variants occur, i.e., the number of variants reduces to K divided by the number of $\hbar$ that fulfill this condition. Note that for $\hbar = 1$, the matching symmetry condition is always fulfilled as it describes the identical symmetry operation. In *MTEX*, unique orientation variants are computed by either of the following two equivalent lines:

```
1 oriC_all = unique(oriP*csP*inv(CRP))
2 oriC_all = variants(CRP,oriP)
```
Code 4

In the specific example of the Burgers OR, the number of unique child variants is reduced from K = 24 to 12 as the two-fold $[110]_\beta$ cubic axis is transformed into the six-fold $[0001]_\alpha$ hexagonal axis.

The problem may now be reversed. Given an OR, ${}_C\mathbf{R}_P$, and a child orientation, ${}_S\mathbf{R}_C$, the possible parent orientation variants, ${}_S\mathbf{R}_P$, are obtained as:

$${}_S\mathbf{R}_P^{\ell} = {}_S\mathbf{R}_C \, \mathbf{S}_C^{\ell} \, {}_C\mathbf{R}_P \tag{3}$$

In combination with the OR, when the crystal symmetries of parent-child phases match, the number of unique variants reduces by the same factor as above. In *MTEX*, the unique variants of the parent phase are computed by either of the following two expressions:

```
1 oriP_all = unique(oriC*csC*CRP)
2 oriP_all = variants(CRP,oriC)
```
Code 5

Thus, for the Burgers OR, the number of unique parent orientation variants reduces from 12 to 6. Here it should be noted that any slight deviation from the Burgers OR breaks the matching symmetry condition and results in 12 parent variants:

```
1 CRP_new = CRP .* orientation.rand(csP,csP,'maxAngle',2*degree);
2 oriP_all = variants(CRP_new,oriC)
```
Code 6

In summary, the true parent orientation of an OR-based phase transformation cannot readily be determined from a single child orientation because of crystal symmetry. A parent grain orientation is determined with a high degree of confidence only when multiple unique and adjacent child grains that originated from the same parent grain via different symmetry operations are identified. This is the primary objective of all parent grain reconstruction algorithms. It follows that if the number of possible orientation variants of the parent orientation increases, higher numbers of unique and adjacent child variants require identification. This is one of the reasons why α′-to-γ parent grain reconstruction in martensitic steels is more challenging than say, α-to-β parent grain reconstruction in Ti or Zr alloys[†]. In this specific comparison, the former involves a transformation of point groups 432-to-432 whereas the latter requires a transformation of point groups 432-to-632.

## 3 Computational approaches to parent grain reconstruction

Computational approaches to parent grain reconstruction presented in the scientific literature to-date are roughly divided into two groups using either, pixel [17,25,26], or grain [16,18–24] -level EBSD map data. The first group of methods claim to be more accurate to local changes in the parent orientation and are apparently superior in identifying annealing twins in austenite and in reconstructing ausformed alloys whereas the second group of methods are said to be computationally more efficient. Considering rapidly increasing EBSD map sizes, the most recently developed parent grain reconstruction algorithms in Refs. [20,22–24] tend to favor the grain-level approach. Consequently, the following paragraphs describe how grain-level parent grain reconstruction methods are implemented in *MTEX*.

A class in *MTEX*, *parentGrainReconstructor*, was designed to contain the methods and properties needed for parent grain reconstruction. The methods enable the selection of different parent grain reconstruction strategies. The properties track the progress of reconstruction and include the OR, the orientation variants, the grain graph and applied weights, the cluster definitions, the reconstructed parent grains, and lists of grain identification numbers that link parent grains with their child grains.

Since all reconstruction algorithms in *MTEX* are grain-level, the parameters chosen during the initial grain reconstruction step are important. Grain reconstruction in *MTEX* is realized by a Voronoi decomposition [29] that is robust against zero solutions (non-indexed or missing pixels) in EBSD maps. In the context of parent grain reconstruction, a threshold angle of say, 3°, is recommended as such a value is: (i) small, (ii) above the orientation noise floor to separate grains, and (iii) avoids large orientation gradients within child grains. Following parent grain reconstruction, neighboring parent grains separated by low-angle boundaries will be merged in any case.

### 3.1 Growth algorithms in partially transformed microstructures

In instances when a significant area fraction of evenly distributed parent phase is retained in partially transformed microstructures or obtained from other reconstruction algorithms, parent grain reconstruction is undertaken by a growth algorithm. The retained parent phase grains represent nuclei that are made to grow into the surrounding child phase. The misorientations at parent-child grain boundaries are compared with the theoretical OR (Eq. (2)) and neighboring parent grains return a vote for the preferred parent orientation of a child grain based on the best fit. Based on the collection of votes from all neighboring parent grains, the best fitting parent orientation is determined via the application of voting metrics. This approach was applied in Refs. [16,18,24,25] to grow the parent phase from previously calculated nuclei.

---

[†] Another reason is that the experimentally determined OR between parent β and child α phases in Ti alloys tends to be closer to a rational OR (in this case, the ideal Burgers OR), which in turn leads to the degeneracy of variants as described above. However, this is not usually the case for α'-to-γ transformation in martensitic steels.

In *MTEX*, the two steps of: (i) computing the votes for a parent orientation using neighboring grains, and (ii) determining the best fitting parent orientation are implemented in the methods *calcGBVotes('P2C')* and *calcParentFromVote*, respectively. Here the option *'P2C'* indicates that only neighboring parent grains to a child grain take part in the vote. The application of these methods is demonstrated in Section 4 via examples. It is re-emphasized that the most crucial parameter for computing the vote is the definition of the threshold angle to identify potential parent-child boundaries.

### 3.2 Nucleation algorithms in fully transformed microstructures

In instances when the parent phase is absent, meaning in fully transformed microstructures, methods that generate nuclei from neighboring child grains may be employed. In their simplest form, nucleation algorithms identify child grains sharing a boundary or triple junction and calculate their disorientation to the OR. The best fitting parent orientation between neighboring child grains is registered as a vote. After collecting the votes from all neighboring child grains, the best fitting parent orientation is determined via the application of additional criteria.

In *MTEX*, the syntax for these two steps is similar to that used for growth algorithms. The methods *calcGBVotes* and *calcTPVotes* compute the votes based on grain boundaries or triple junctions, respectively. The determination of the parent orientations from the votes is done by the method *calcParentFromVote*.

The criterion for parent reconstruction in the initial algorithms by Humbert et al. [15] on Ti and Zr alloys and Cayron et al. [16] on steel required the identification of three child variants belonging to a common parent grain. The latter subsequently applied a growth algorithm to finalize the parent reconstruction. The criterion applied by Germain et al. [18] comprised an iterative procedure that graphically searched neighboring child grains with low disorientation to a given OR and computed the best fitting common parent orientation for these grains. Following the nucleation stage, a growth algorithm was applied to reconstruct the parent orientation for the remaining child grains. Most grain-level algorithms described in Refs. [21–24] are based on this approach and apply various adjustments to improve specific parent grain reconstruction scenarios.

### 3.3 Graph clustering algorithms

While nucleation and growth algorithms begin reconstruction locally and evolve iteratively to the full map, graph clustering algorithms work on a global map scale right from the start. These algorithms assign an OR probability value as a weight to the edges of grain graphs. The OR probability is a parameter derived from the disorientation between grain misorientations and the OR. Subsequently, a graph clustering algorithm is applied to the grain graph by clustering together all child grains that are likely to belong to the same parent grain. The third step fits parent orientations to each of these clusters. Graph clustering reconstruction algorithms were previously proposed by Gomes et al. [19] and Nyyssönen et al. [20].

In *MTEX*, these three steps are applied in the methods *calcGraph*, *clusterGraph* and *calcParentFromGraph*, respectively.

## 4 Example applications of *MTEX* to parent grain reconstruction

In the following section, the syntax and functionality of parent grain reconstruction in *MTEX* is demonstrated for three example alloys that undergo well-known phase transformations namely, **α′**-to-**γ** in a lath martensitic steel, **α**-to-**β** in a Ti alloy, and a two-step transformation, **α′**-to-**ε**-to-**γ**, in a twinning and transformation -induced plasticity steel. Some of the more advanced plots are created using *ORTools* [28].

### 4.1 α′-to-γ reconstruction in a lath martensitic steel

The EBSD map data is courtesy of Nyyssönen et al. [20]. The microstructure, shown in Figure 1, consists of lath martensite (α′) and 27% unindexed points. The script used to reconstruct the parent γ grains from a child α′ microstructure is available via Ref. [30].

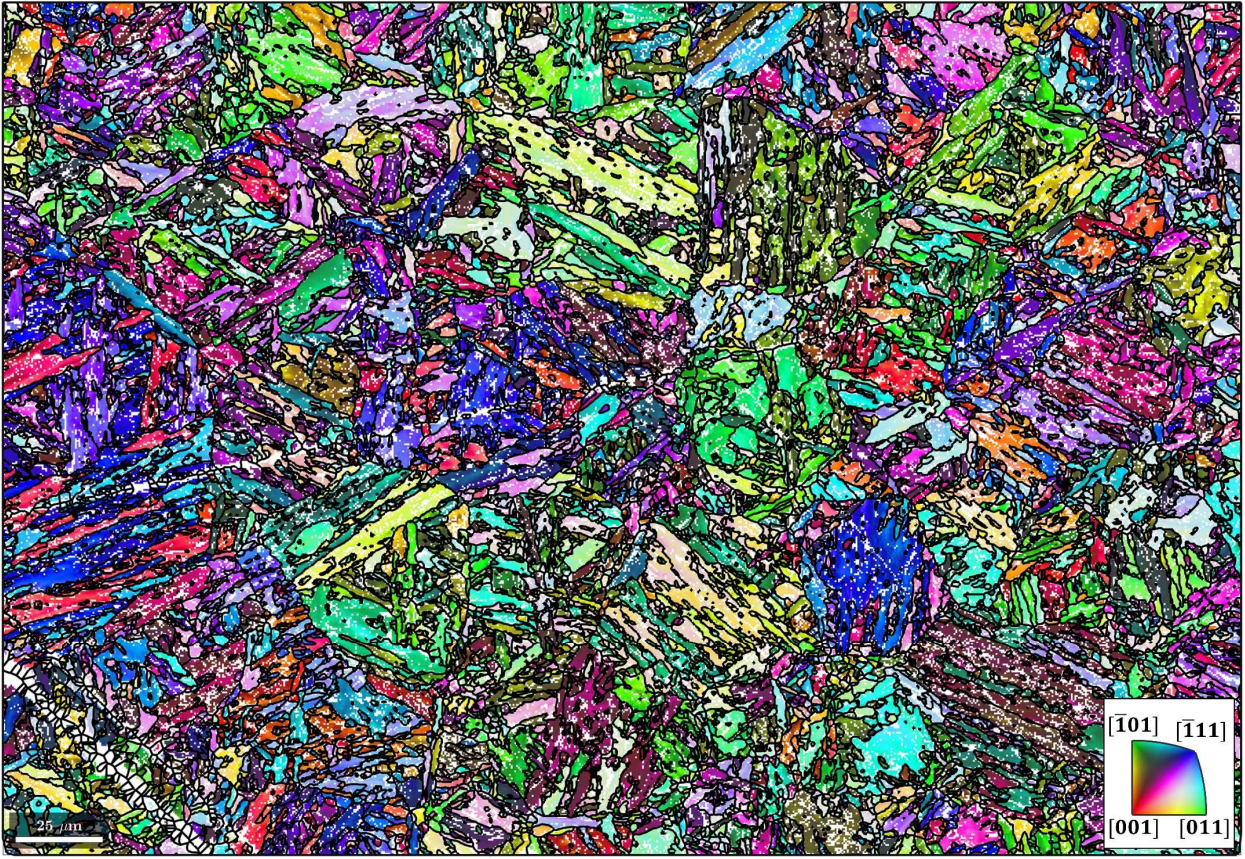

**Figure 1:** Inverse pole figure map of lath martensite. The martensite grains are identified by a misorientation threshold of 3°. Grain boundaries are in black and zero solutions are in white.

### 4.1.1 Irrational OR determination from EBSD map data

Before parent grain reconstruction can begin, the irrational OR from EBSD map data, which usually lies somewhere in-between the rational K-S and Nishiyama-Wassermann ORs in lath martensite, needs to be determined. In this example, the accurate determination of the irrational OR from EBSD map data is necessary to increase the success rate of parent variant indexing, which in turn may also enable the detection of twinned parent grains, if any [17].

In *MTEX*, parent grain reconstruction begins by constructing an object, *job*, from the *parentGrainReconstructor* class by supplying the *ebsd* data and the computed *grains* (Code 7.1). An initial guess for the OR is provided by assigning the K-S OR as a misorientation to the property, *p2c* (Code 7.2). The method, *calcParent2Child* (Code 7.3), then determines the irrational OR from EBSD map data via iterative refinement:

```
1 job = parentGrainReconstructor(ebsd, grains)
2 job.p2c = orientation.KurdjumovSachs(csAlpha,csGamma)
3 job.calcParent2Child
```
Code 7

Accurate irrational OR determination from EBSD map data is a necessary requirement for successful parent grain reconstruction. This is especially the case when an OR contains many orientation variants and/or deviates significantly from the closest rational OR, as shown in this example [17].

The method *calcParent2Child* is an improvement of the iterative OR refinement procedure presented in Ref. [31]. The refinement method iteratively identifies boundary misorientations that reasonably agree with the current best guess of the OR and refines the best guess in each iteration by taking the mean of the identified boundary misorientations. Once the refinement procedure has converged, the property *p2c* is updated with the OR that gives the best fit to the misorientations in the microstructure.

In Figure 2a, the disorientations between martensitic grain misorientations and the misorientations of the initial K-S and refined ORs are shown. It is evident that the disorientation was minimized by iterative OR refinement. It is also clear that no OR exists that reduces the disorientation for all grains to zero in the present microstructure. Adapting the OR to reduce the disorientation for some grains would inevitably lead to larger disorientations for other grains. The (001) pole figure of the 24 martensitic variants of the refined OR is given in Figure 2b.

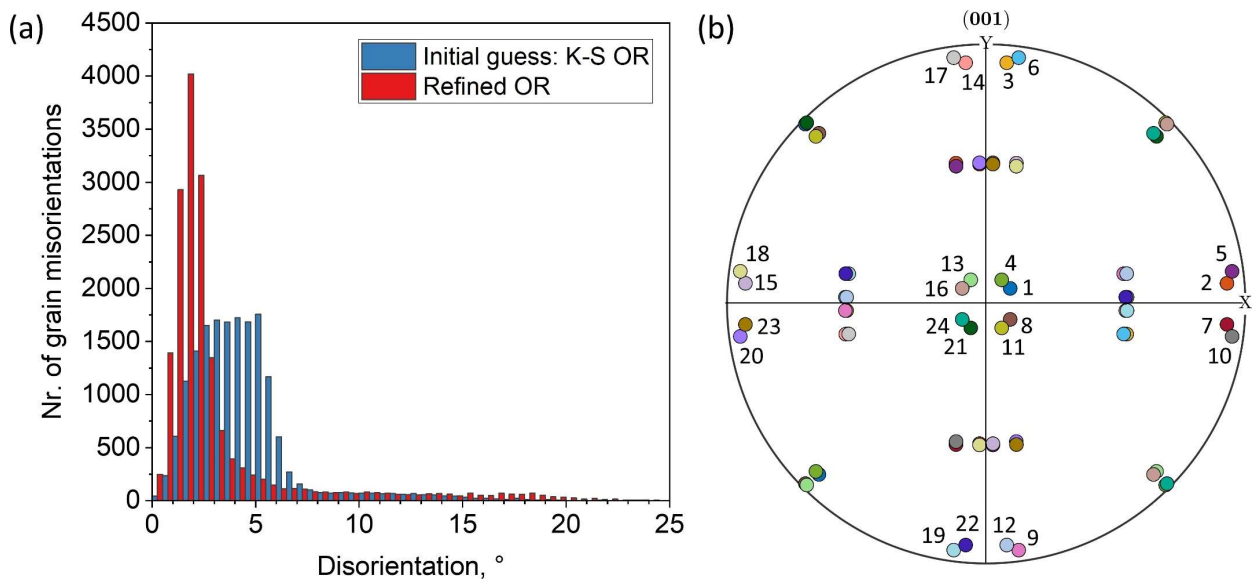

Figure 2: (a) Disorientation histogram between martensitic grain misorientations and the misorientations of the K-S and refined ORs. (b) {001} pole figure showing the 24 martensitic variants of the refined OR.

### 4.1.2 Building and clustering the weighted grain graph

The disorientation between martensitic grain misorientations and the refined OR (Figure 2b) are plotted by color-coding the martensitic boundaries in Figure 3 with a threshold of 5°. It is obvious that the network of boundaries with disorientations >5° corresponds to prior austenite grain boundaries. Although short segments along such boundaries have disorientations <5°, the latter are, on balance, likely to be prior austenite grain boundaries based on their connectivity to similar boundaries with overall disorientations >5°. Keeping the above in mind and considering the many orientation variants in this example, the low disorientation across the short boundary segments of prior austenite grain boundaries is ascribed to coincidence.

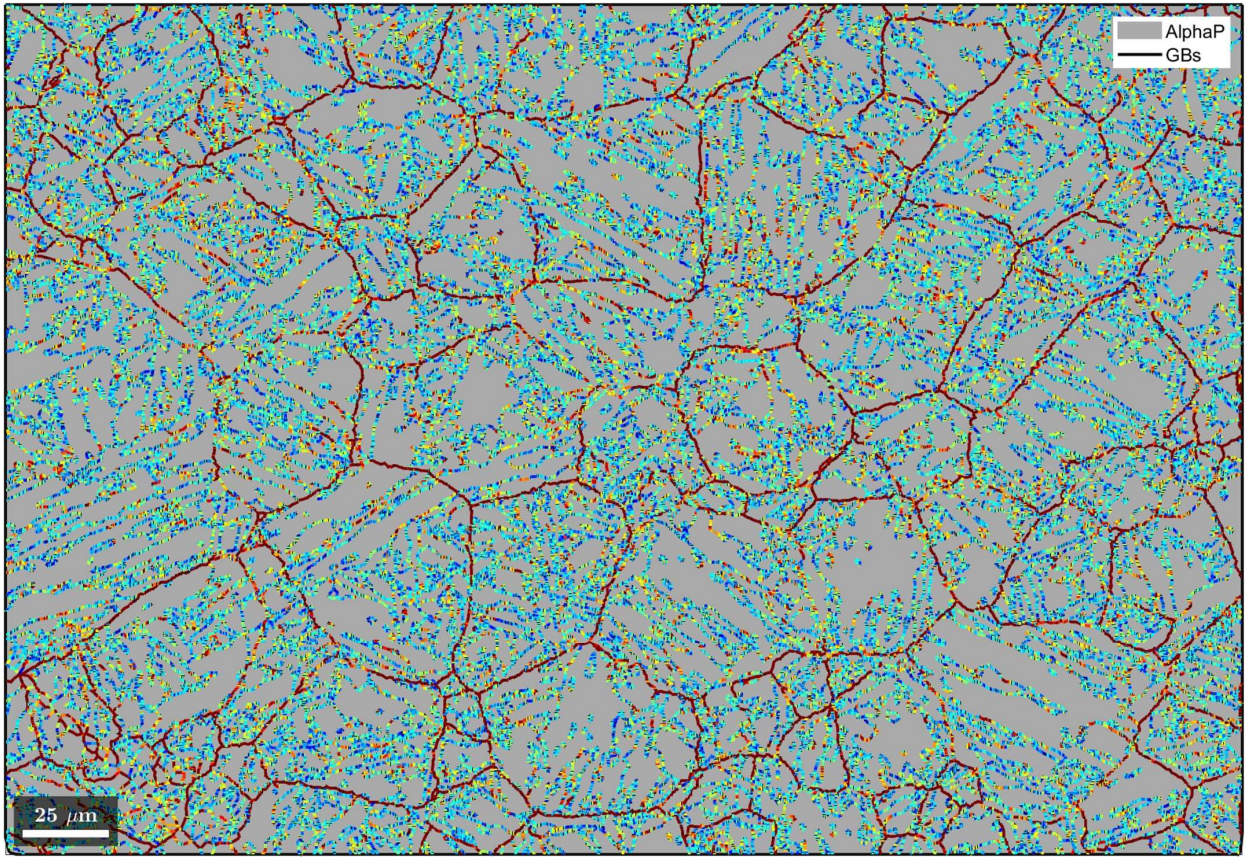

**Figure 3: Distribution of the disorientation histogram in Figure 2b visualized by color-coding the martensitic boundaries with a threshold of 5°.**

Using the methods described in Section 3.3, a graph of martensitic grains is subsequently built. Neighboring grains are connected by edges that are weighted by the probabilities of them belonging to the same parent grain. The probability is derived from the disorientation shown in Figure 3 and is expressed by a cumulative Gaussian distribution with a given mean and standard deviation. In *MTEX*, this functionality is integrated into the method, *calcGraph* (Code 8.1).

After building the graph, a clustering algorithm is applied to identify clusters of strongly connected grains according to the above calculated probability by the method, *clusterGraph* (Code 8.2). By default, this method features a Markov clustering (MCL) algorithm, which simulates a random walk across nodes that connect neighboring grains. MCL is an attractive choice for the current application as it is: (i) an unsupervised algorithm, (ii) computationally efficient, and (iii) resistant to noise [19]. The OR probability assigned to the nodes is equivalent to the probability with which the MCL algorithm walks along the different nodes.‡ With each iteration of random walk, nodes that connect grains belonging to the same parent orientation are gradually strengthened whereas nodes connecting the grains that do not belong to the same parent orientation are gradually cut off. The resulting clusters are depicted by black boundaries overlaid on a semi-transparent inverse pole figure martensite map in Figure 4. Most of the prior austenite grains (depicted by the predominantly red outlines in Figure 3) are divided into several clusters, and most clusters contain different martensitic variants.

```
1 job.calcGraph
2 job.clusterGraph
```
Code 8

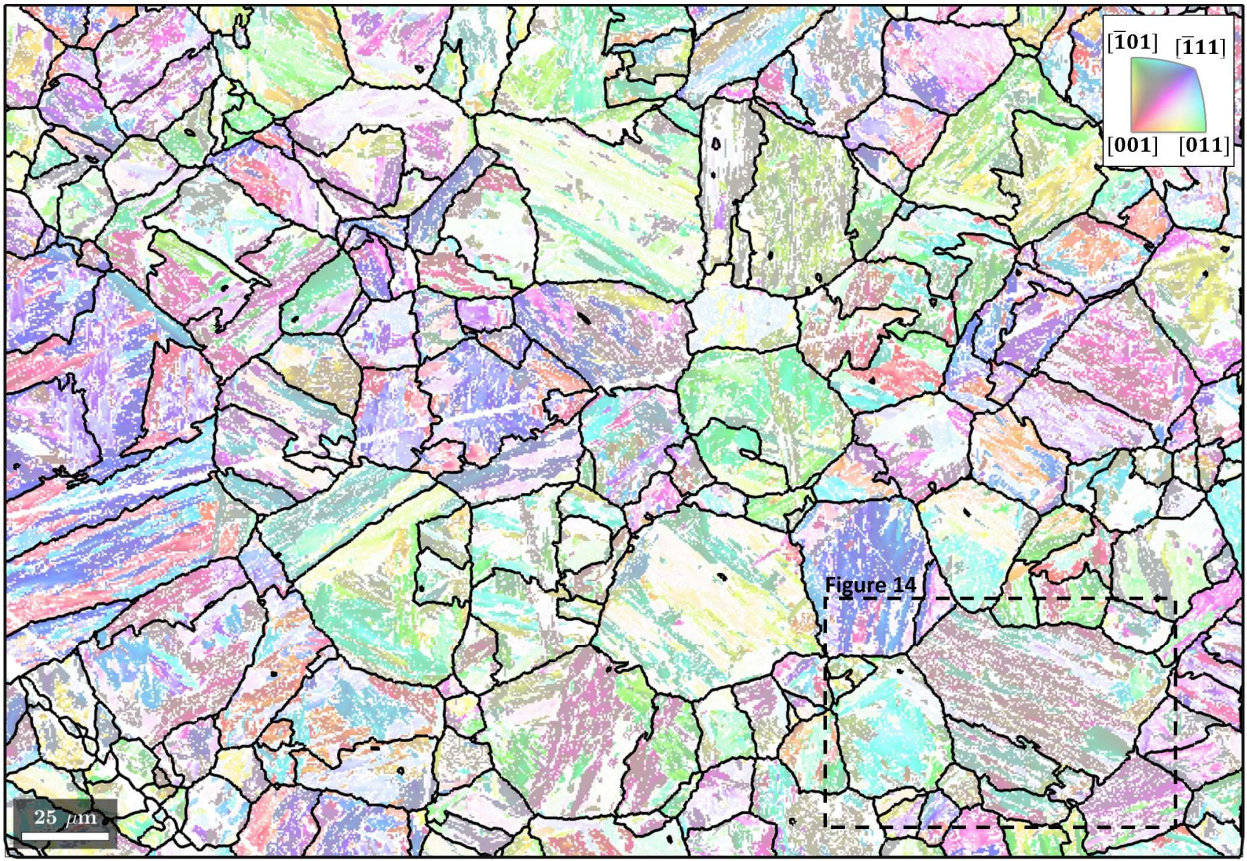

**Figure 4: Clusters of martensite grains outlined in black and overlaid on the inverse pole figure map from Figure 1.**

---

‡ It could be equivalently stated that the MCL algorithm walks are random along the nodes by a fixed distance per iteration and that the path lengths of the nodes are the inverse of the probabilities assigned to the nodes.

### 4.1.3 Reconstructing parent grain microstructures from grain clusters

After identifying the clusters of martensitic grains that are likely to belong to the same austenite grain with the method, *clusterGraph*, they are transformed to parent orientations by applying the method, *calcParentFromGraph*. The calculation consists of two steps as follows:

i. All possible parent orientations are calculated by applying the inverse OR to each child grain orientation in the cluster as per Eq. (3). A common parent orientation is computed by minimizing the overall disorientation to a possible parent grain orientation that is common or close to all child grain orientations in the cluster. The area of the child grains is used as the weight in this fitting procedure.
ii. The parent orientation of each child grain in a cluster is determined by calculating the parent orientation with the least disorientation to the common parent orientation of the cluster by applying the OR.

The procedure produces the reconstructed clusters in Figure 5. The regions previously identified with <5° disorientation to the OR in Figure 3 are regions with a common parent austenite orientation.

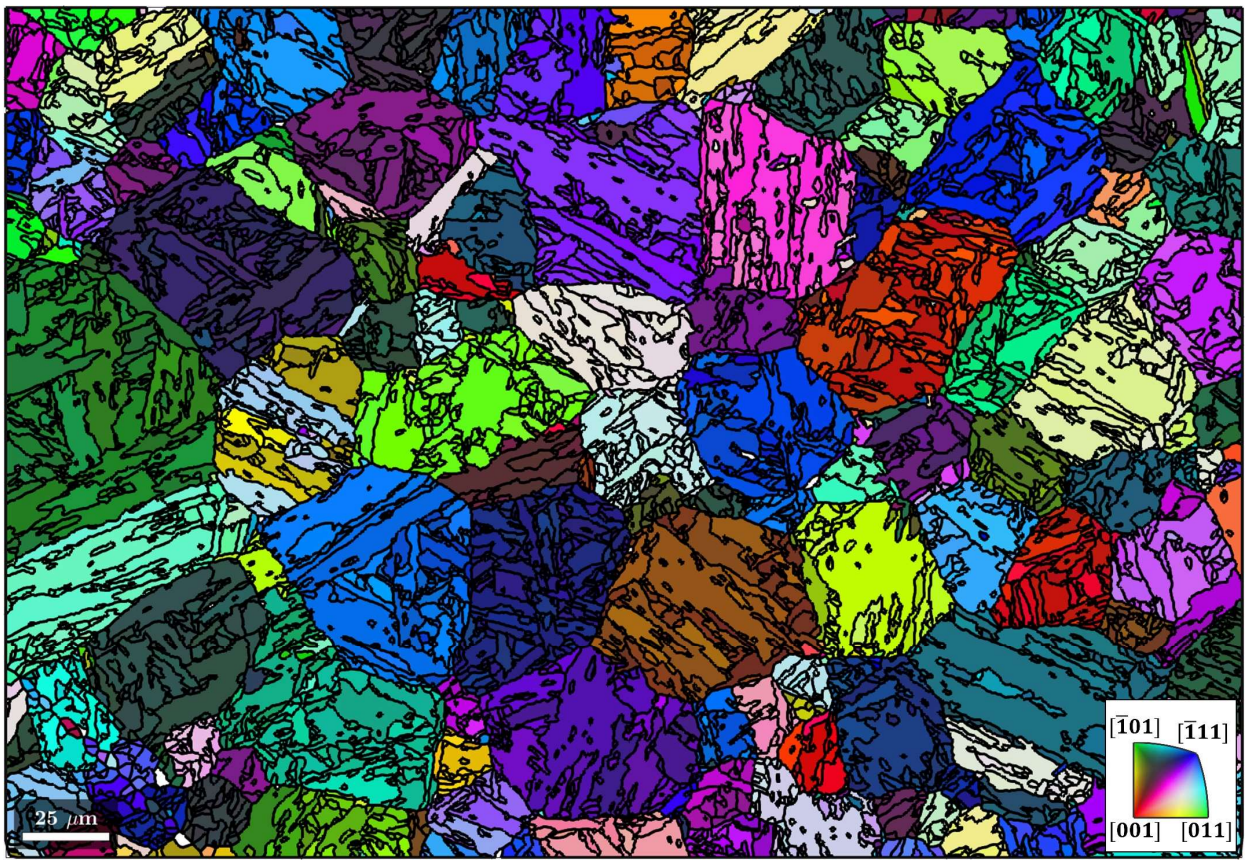

**Figure 5: Reconstructed parent microstructure from child grain clusters shown in Figure 4 with the method, *calcParentFromGraph*.**

### 4.1.4 Evaluation and local reversion of the reconstruction

After calculating the common parent orientation of each cluster and the parent orientation of each child grain in the cluster, the disorientation between them may be evaluated. This disorientation is plotted by applying a 5° threshold in Figure 6.

Martensite grains with a high disorientation are likely to have been assigned to the wrong cluster in the above procedure. Very small clusters are also likely to yield an uncertain parent orientation. Consequently, the disorientation and cluster size criteria (and any others if defined by a user) are applied to revert such poorly reconstructed martensite grains by the method, *revert*:

```
1 job.revert(job.grains.fit > 5*degree)
2 job.revert(job.grains.clusterSize < 15*degree)
```
Code 9

It follows that after reversion, the orientations of the remaining reconstructed martensite grains have a higher likelihood of being actual parent grains (see Figure 7).

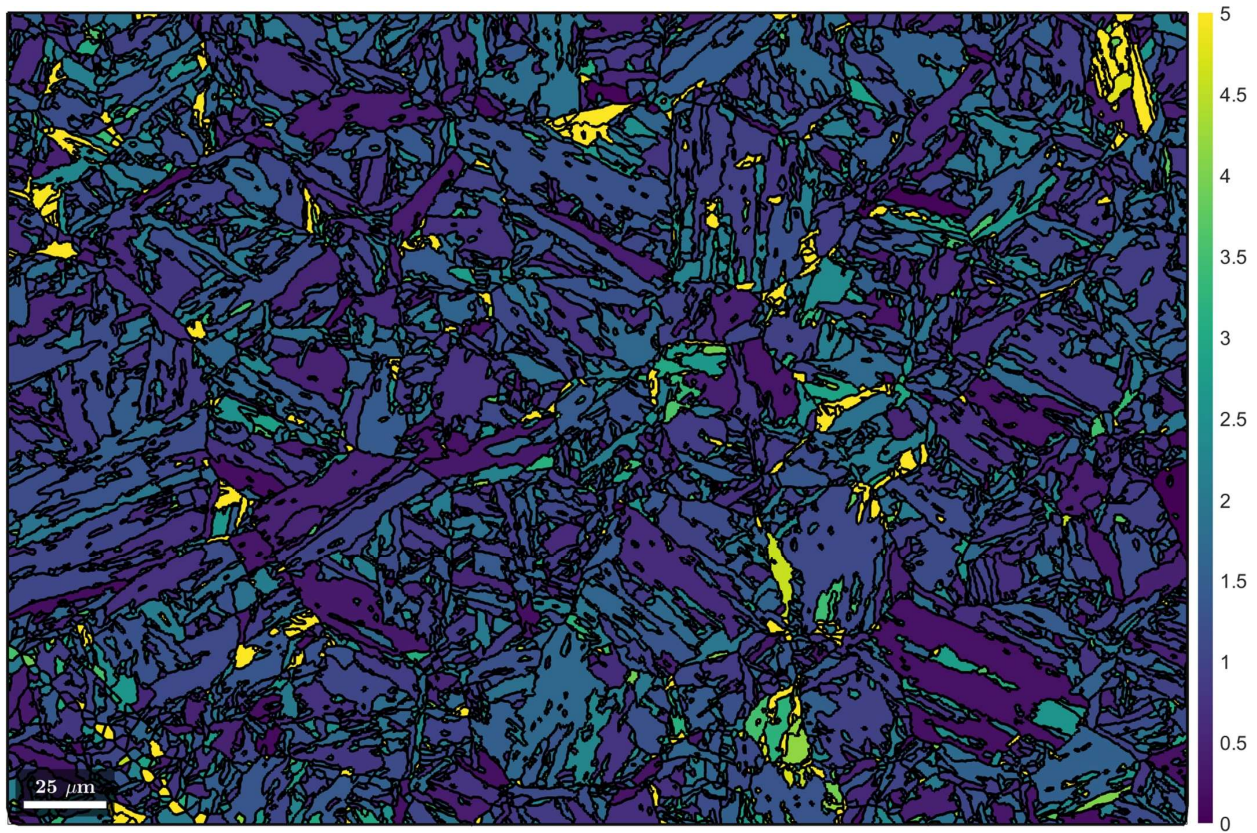

**Figure 6: Disorientation between the common parent orientation of each cluster and the parent orientation of each child grain in the cluster with a threshold of 5°.**

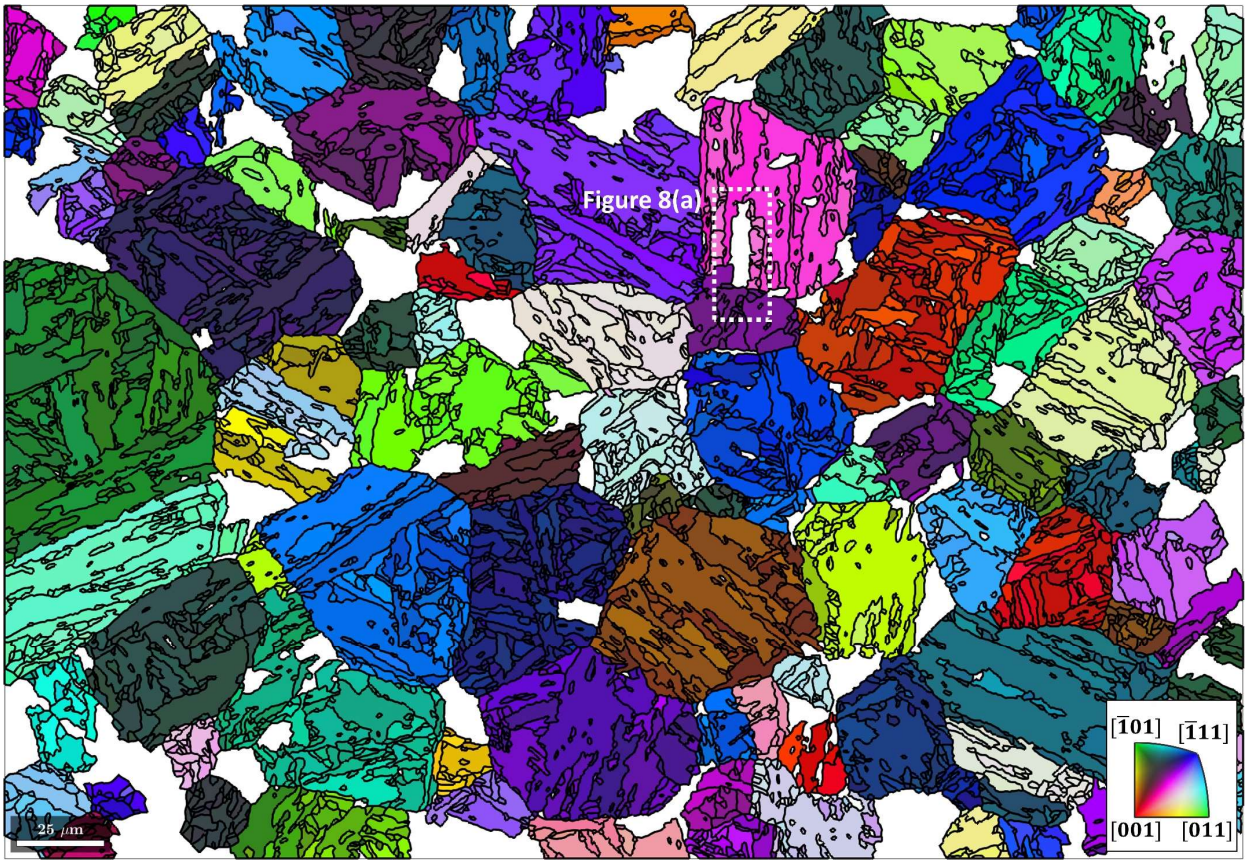

**Figure 7: Remaining reconstructed parent microstructure after reverting certain grains with Code 9. The reverted regions are in white.**

### 4.1.5 Reconstructing the parent grain microstructure by a growth algorithm

Since the reconstruction of a significant fraction of martensite grains was unsuccessful within the given confidence criteria (Figure 7), the remaining reconstructed parent phase serves as nuclei in a growth algorithm (see Section 3.1) within the reverted regions:

```
1 job.calcGBVotes('p2c', 'threshold', 2.5*degree);
2 job.calcParentFromVote
```
Code 10

Figure 8a is an example of this algorithm at work in a local region defined by the dotted white rectangle in Figure 7. In this example, the white area corresponds to martensite grains that have two neighboring parent grains and have not yet been reconstructed. In the method, *calcGBVotes*, the boundary misorientations of all possible parent orientations of a child grain with neighboring parent grains are computed and voting probabilities are assigned. The threshold angle, 2.5°, marks the misorientation angle between a neighboring parent orientation and the reconstructed parent orientation of the child grain at which the probability is 50 %. After three iterations of Code 10, the reconstructed parent grain microstructure in Figure 8b is obtained.

Although the microstructure is not fully reconstructed, the reconstructed areas have a high confidence. Depending on the microstructure, parent grain reconstruction may be continued with lower confidence criteria. It is apparent that regions located at or near the vertical and horizontal edges of the EBSD map were not reconstructed. This could be ascribed to the lack of neighboring grains and unique child orientation variants in these regions.

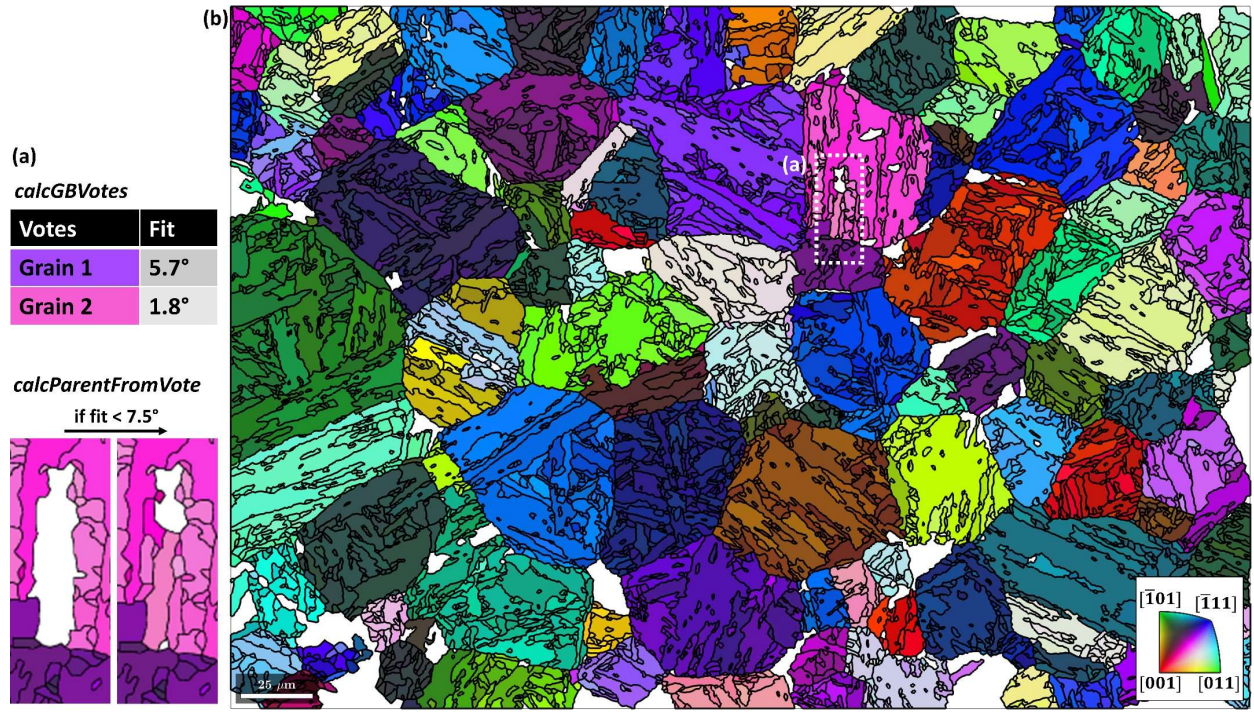

Figure 8: (a) A local example of the growth algorithm at work. (b) Reconstructed parent grains after three iterations of the growth algorithm.

### 4.1.6 Cleaning the parent grain microstructure and reconstructing the EBSD data

In Figure 8 the martensite grains are reconstructed to largely similar parent orientations within any prior austenite grain. The misorientation between these fragmented grains is used to merge them to prior austenite grains by calling the method, *mergeSimilar*, with a threshold for the maximum allowed misorientation angle between neighbouring grains (Code 11.1). Subsequently, the method, *mergeInclusions*, merges small grains within a specified maximum area (Code 11.2).

```
1 job.mergeSimilar('threshold',7.5*degree);
2 job.mergeInclusions('maxSize',50);
```
Code 11

In this way, child grain clusters containing common parent orientations (Figure 8) are transformed into parent grains (Figure 9). A distinguishing feature of these methods is that the merging process is tracked by the property, *mergeId*. This property enables users to list the child grains belonging to a particular parent grain and is crucial for subsequent variant analysis.

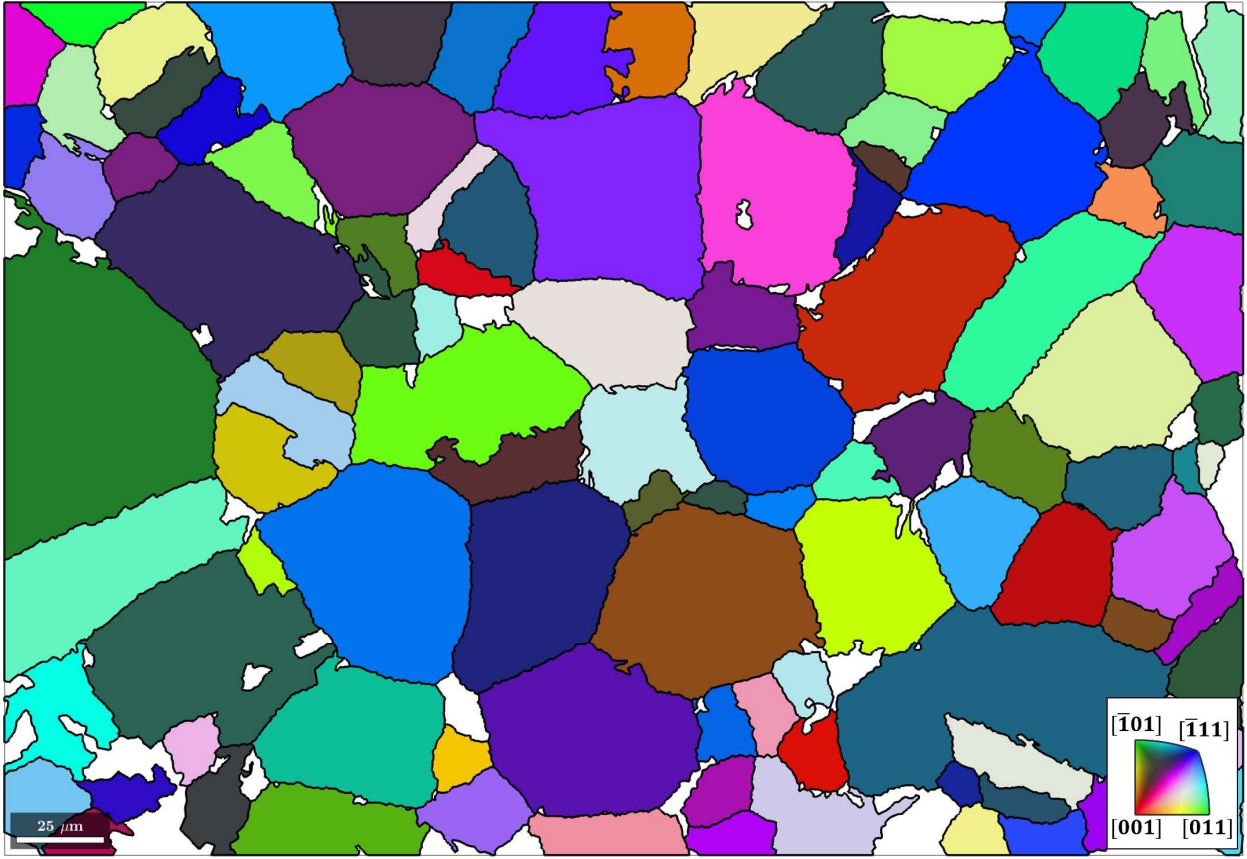

**Figure 9: Reconstructed parent grain microstructure after applying the cleaning steps in Code 11.**

In a final step, the EBSD data of the parent phase is reconstructed from the EBSD data of the child phase by the method, *calcParentEBSD*. Here the grain-level record of the particular parent orientation variant reconstructed for each child grain is applied to the child EBSD data (Figure 1) and the OR. . The resulting parent EBSD data is shown in Figure 10 with the prior austenite grain boundaries overlaid in black.

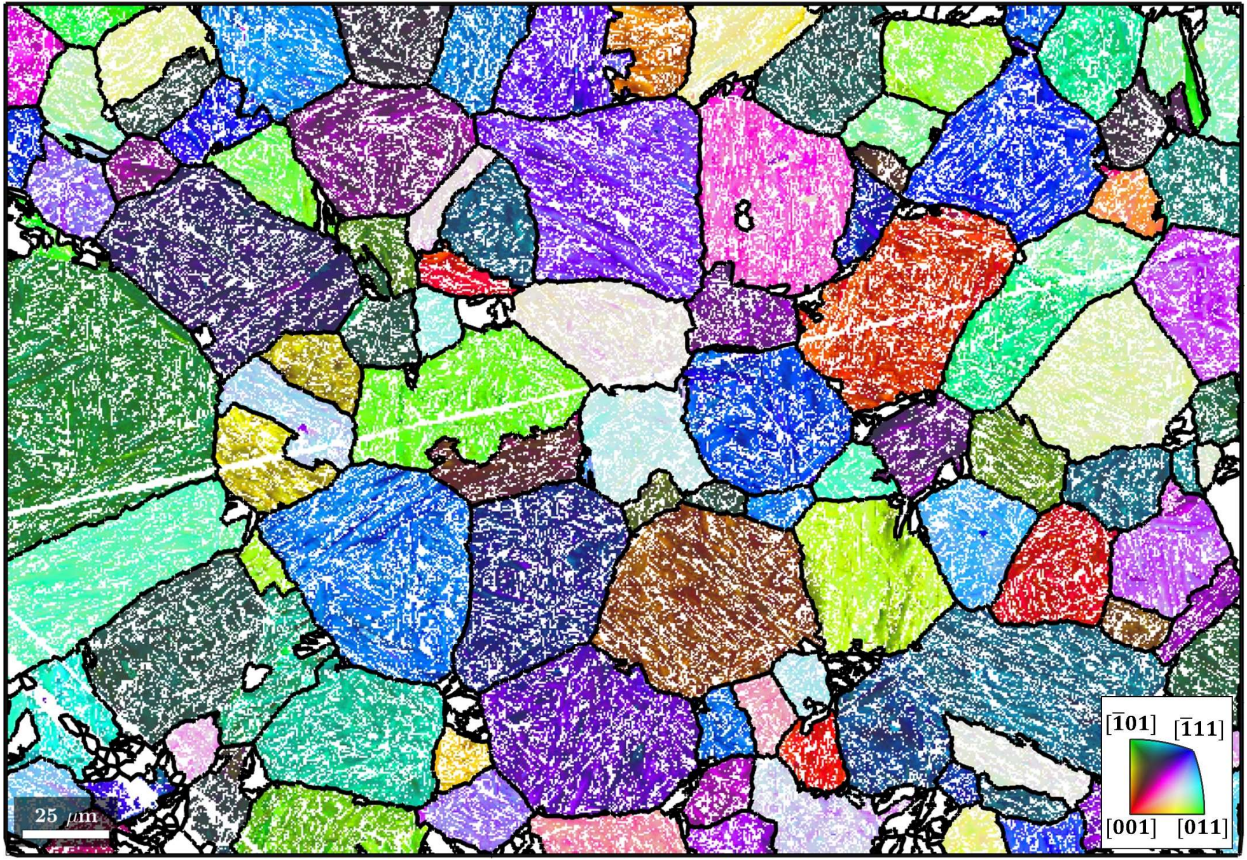

Figure 10: Reconstructed parent EBSD data with the parent grain boundaries in black.

### 4.2   α-to-β reconstruction in a Ti alloy

The EBSD map data is courtesy of Susanne Hemes, Access e.V. The initial microstructure consists of 94.5% α and 0.2% β phases and 5.3% unindexed points and is shown in Figure 11a. The script used to reconstruct the parent β grains from a child α microstructure is available via Ref. [30].

Similar to Section 4.1, the object, *job*, is constructed from the *parentGrainReconstructor* class and the OR is initialized as the Burgers OR [32], as shown in Section 0 while summarizing the theory of parent grain reconstruction:

```
1 job = parentGrainReconstructor(ebsd, grains)
2 job.p2c = orientation.Burgers
```
Code 12

The rotation axes of α − α boundary misorientation pairs are plotted in Figure 11b along with the six ideal Burgers OR variant pairs shown as white circles =. The misorientation axes of α − α boundary pairs are color-coded based on their disorientation to the Burgers OR. The α − α boundary pairs have low disorientation values and their misorientation axes are close to that of the ideal Burgers OR as well. The few data points with high disorientation and/or different misorientation axes likely conform to a different OR and/or belong to prior $\beta$ boundaries. The conformity of the microstructure to the ideal Burgers OR indicates that OR refinement is not needed. Since the number of variants are few and distinctly defined, and the grain morphology contains a large density of triple points, a triple-point parent reconstruction strategy may be applied:

```
1 job.calcTPVotes('minFit',2.5*degree,'maxFit',5*degree);
2 job.calcParentFromVote('minProb',0.7);
```
Code 13

The nucleation method, *calcTPVotes* (see Section 3.2), is applied to identify triple points and determine the best fits for a parent orientation between three grains (Code 13.1). Here only triple junctions for which the best fit is below 2.5° and the second best fit is above 5° are considered. The above method is similar to the method, *calcGBVotes*, used in the growth algorithm of the previous example. Following this, the method, *calcParentFromVote*, is applied to reconstruct all grain clusters with a probability of at least 0.7. The method returns the reconstructed microstructure in Figure 11c. A single iteration of the growth algorithm (Code 10) and reconstructing the parent EBSD data yields the final parent grain microstructure shown in Figure 11d.

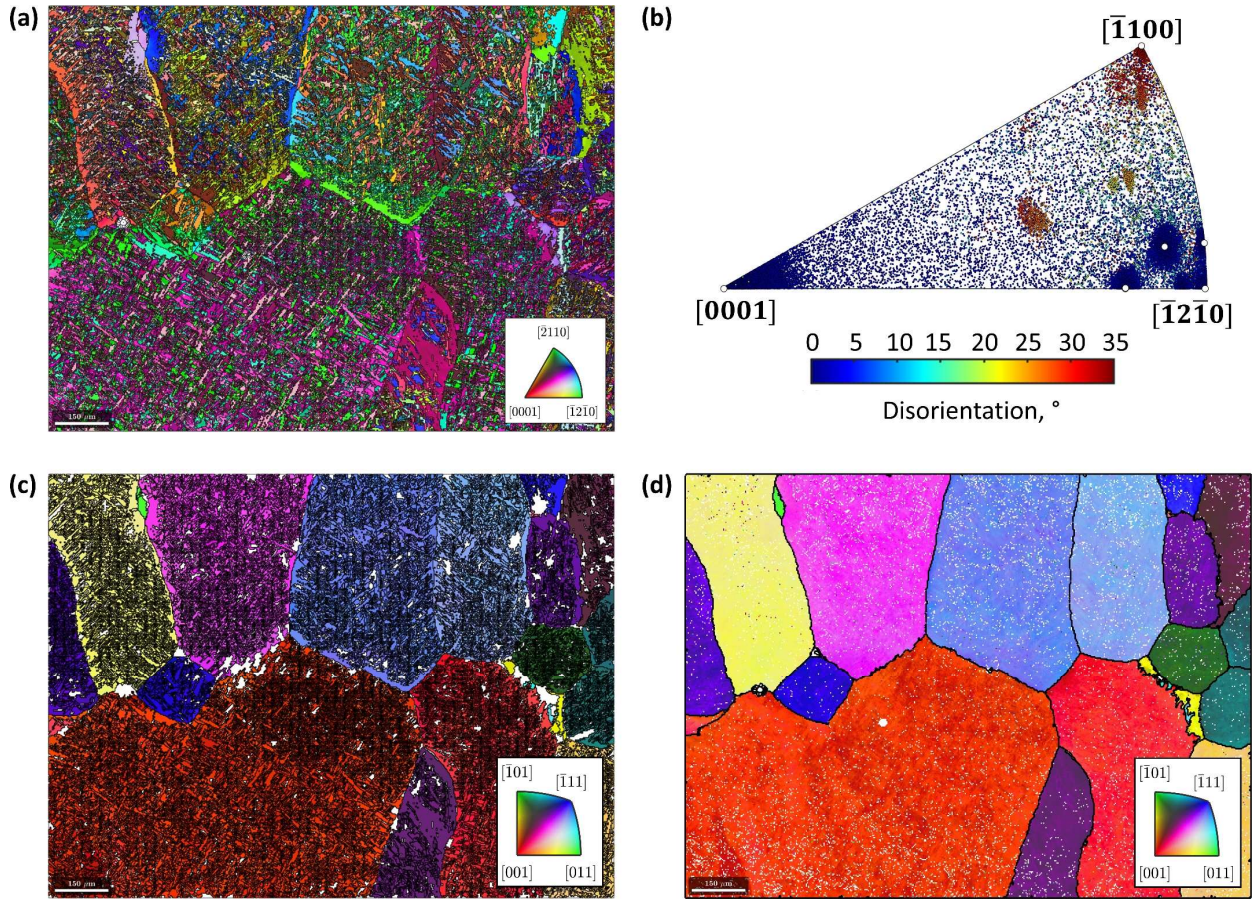

Figure 11: (a) IPF map of the child α phase. (b) IPF of the misorientation axes between α − α boundary pairs color-coded according to their disorientation to the ideal Burgers OR along with the six ideal Burgers OR variant pairs shown as white circles (c) Parent grain microstructure reconstructed from α triple points. (d) Reconstructed parent EBSD data after applying α triple point and growth algorithms. The black lines are the parent grain boundaries.

### 4.3 α′-to-ε-to-γ reconstruction in a twinning and transformation -induced plasticity steel

The EBSD map data is courtesy of Pramanik et al. [33]. The script used to reconstruct the parent ε and γ grains is available via Ref. [30]. As shown in Figure 12a, the initial microstructure consists of 56% face-centered cubic parent γ, 26% hexagonal close-packed ε and body-centered cubic 18% α′. ε and α′ formed during quenching after prior hot-rolling as well as the partial transformation of γ-to-ε, γ-to-α′, and ε-to-α′ via the Shoji-Nishiyama, K-S and Burgers ORs, respectively, during subsequent cold-rolling to 10% thickness reduction.

The orientations of each phase and the computed grain boundaries are shown in Figure 12c-12e. The modular setup of the *MTEX* parent grain reconstruction algorithm allows for the complete reconstruction of parent γ via a two-step process in a single workflow. Since sufficient parent-child boundaries are present for both martensite transformations, the *ORTools* function, *peakFitORs*, is used to determine the ORs between γ-ε and ε-α′ by fitting the parent-child boundary misorientation angle distribution (Figure 12b). The workflow is the same as the one described in Section 0 for the parent grain reconstruction of lath martensite and comprises the sequential application of clustering, reconstruction, reverting bad fits, growth, cleaning and calculation of EBSD data. The workflow is first applied to reconstruct all α′ grains to ε (Figure 12f) and subsequently, to reconstruct all ε grains to γ (Figure 12g). Throughout the entire workflow, the IDs of the child grains are stored in the object, *parentGrainreconstructor*. In this way, an advanced transformation graph over two orders of transformation is constructed by linking the grain identification numbers of all child grains to their parent grain(s). Thus, an α′ grain transformed from an intermediate ε grain is uniquely identified and linked together. Concurrently, both grains are also linked to the single reconstructed parent γ grain from which they transformed. Since the grain identification number links all parent grains to their child grains, it also enables second order variant analysis.

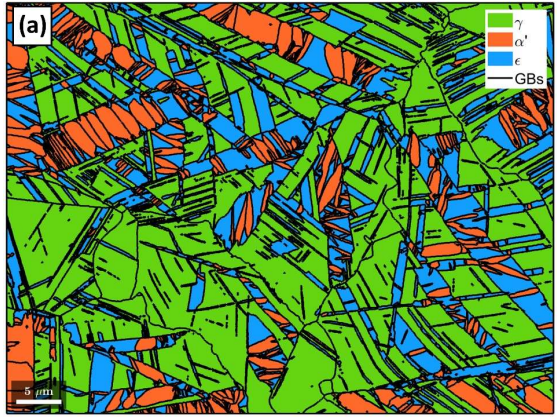
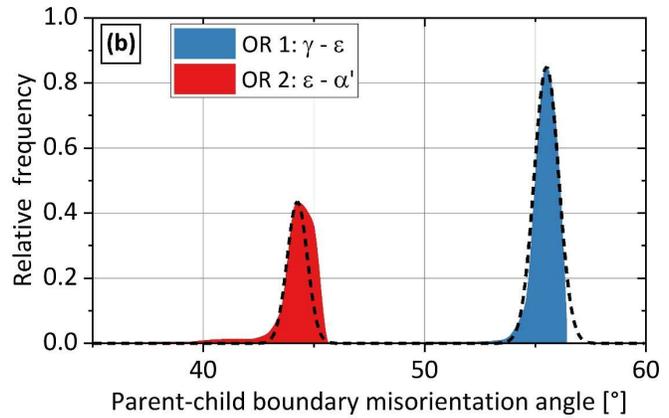

**Transformation ⇒** $\gamma \xrightarrow{OR_1} \varepsilon$ $\varepsilon \xrightarrow{OR_2} \alpha'$

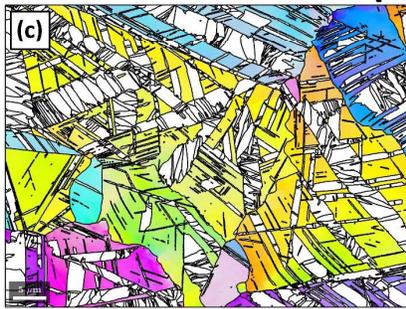
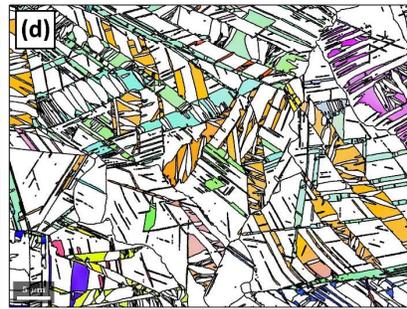
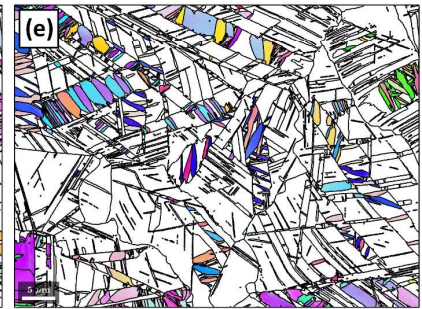

**Reconstruction ⇒** $\alpha' \xrightarrow{OR_2^{-1}} \varepsilon$ $\varepsilon \xrightarrow{OR_1^{-1}} \gamma$

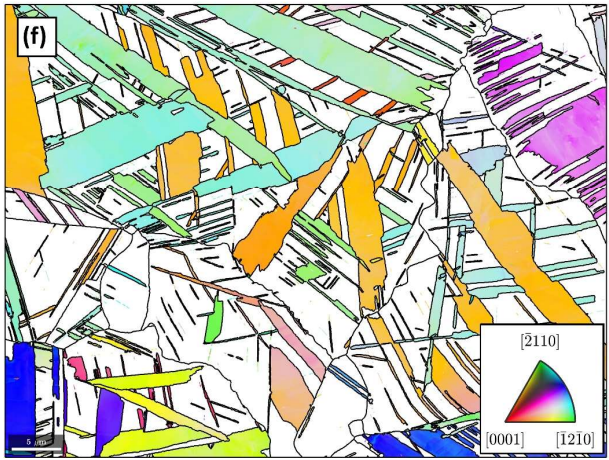
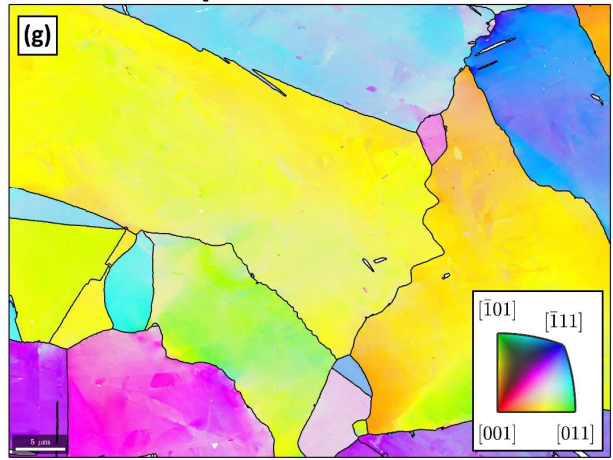

**Figure 12: Parent grain reconstruction of the two-stage martensite transformation γ-to-ε-to-α′.** The phase map (a) shows the initial phase distribution. The ORTools function, peakFitORs, is used in (b) to fit both ORs based on the parent-child misorientation angle distribution. The inverse pole figure maps of (c) γ, (d) ε and (e) α′ show the initial grain orientations. The sequential reconstruction of (f) α′-to-ε and (g) reconstructed + retained ε to γ is carried out in a single workflow. The colors in (c) and (e) are as per the inset stereogram in (g) whereas the colors in (d) are as per the stereogram in (f).

# 5 Discussion

## 5.1 Highlights of the *MTEX* implementation

The examples of parent grain reconstruction in Section 4 demonstrate the versatility of the newly integrated functionalities for parent grain reconstruction in *MTEX*. The main advantage of the present approach lies in the modularization of the reconstruction process as defined by the class, *parentGrainReconstructor*, which contains the essential methods and properties for parent grain reconstruction (see Section 3). This approach enables the creation of individual workflows and reconstruction strategies for different types of transformation microstructures. Additional ancillary methods such as the local reversion of reconstruction and the merging of similar grains round-off the core functionality. Therefore, the above approach is an ideal trade-off between automation and versatility.

## 5.2 Computational performance

The code has been optimized for speed by using efficient vectorized expressions in MATLAB. The α'-to-γ transformation in Section 0 was timed as follows. The EBSD data contains 486 × 707 pixels and reconstructed 7002 martensitic grains. The example was calculated on a contemporary office laptop on a single Intel® Core™ i7-8650U processor core with a 1.9 GHz processor base frequency. The refinement of the OR by the method, *calcParent2Child*, took 14 s. The execution of the methods, *calcGraph, clusterGraph* and *calcParentFromGraph*, for the first part of the parent reconstruction took 37 s. Three loops of the growth algorithm (Code 10) took a further 2 s.

In the α-to-β transformation in Section 4.2, the map comprised 384 × 512 pixels and reconstructed 49,666 grains. The triple-point based reconstruction (Code 13) took just under 5 s and a single iteration of the growth algorithm (Code 10) took an additional 1 s on the same computer setup.

## 5.3 Orientation variant analysis

Reconstructing the parent grain microstructure and parent EBSD data is a prerequisite for in-depth orientation variant analysis. The tools to compute orientation variant and packet identities are implemented in the method, *calcVariants*. With a few more additional lines of MTEX code, plots associated with variant analysis can be produced. However, the add-on *ORTools* [28] already features several pre-written functions to create publication-ready plots associated with variant analysis.

Figure 13 is an example of variant analysis using *ORTools* on the reconstructed parent γ microstructure in Section 0. By default, *MTEX* assigns the convention for packet and variant numbering established by Morito et al. [34] whenever a transformation between two cubic symmetries is detected. In Figure 13a, the EBSD data of lath martensite is colored according to packet identification numbers that delineate martensitic variants formed from the same habit plane. In this example, the four habit planes are: $(111)_\gamma$, $(1\bar{1}1)_\gamma$, $(\bar{1}11)_\gamma$, or $(11\bar{1})_\gamma$. *ORTools* enables the investigation of individual reconstructed parent grains by an interactive function, *grainClick*. Some of the obtained plots are shown in Figure 13b to 13f. The variant map in Figure 13b shows the variant identification numbers of the EBSD data and the martensite grain boundaries. It is evident that most martensite grains, which represent martensitic blocks, contain pairs of two different variants. The pairing is according to the common Bain groups within a packet and is commonly observed in lath martensitic steel [34,35]. For instance, for packet 1, the pairing is of the type V1-V4, V2-V5 and V3-V6. Equivalent pairing may also be denoted for the three other packets. Figure 13c is the $(001)_\gamma$ pole figure of the theoretically predicted martensite variant orientations based on the parent $\gamma$ mean grain orientation and the OR. Figure 13d shows excellent agreement between the predicted and observed variant orientations. Finally, the Figure 13e and Figure 13f show the distributions of the variant and packet identification numbers within the parent grain, respectively.

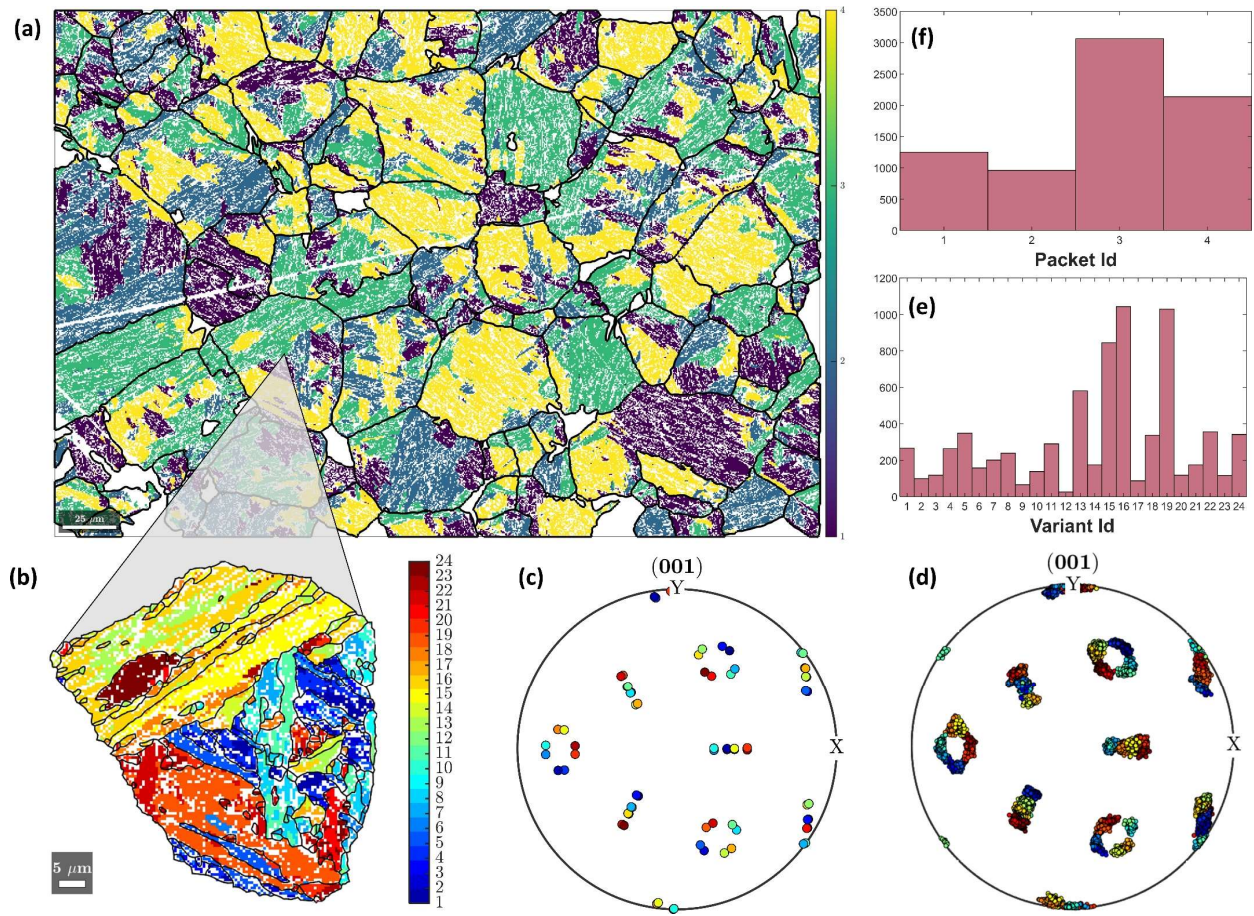

**Figure 13: Example of variant analysis on the reconstructed lath martensite steel microstructure from Section 0 using the add-on function library, *ORTools* [28] (a) EBSD map showing the packet Ids of martensite and grain boundaries of the reconstructed parent grains. (b) EBSD map showing the variant Ids of martensite and boundaries of the martensite grains. $(001)_\gamma$ pole figures of the (c) predicted, and (d) observed martensite variants of the highlighted parent γ grain in (b). Area fractions of the (e) variants and (f) packets of the highlighted parent γ grain in (b).**

## 5.4 Reconstruction of austenite annealing twins in martensitic steel

While the present study highlights the versatility of the new framework for parent grain reconstruction in *MTEX*, it is also appropriate to discuss how the new methods approach the common problem of reconstructing annealing twins in parent austenite from child martensite grains in steel microstructures [17].

The problem is demonstrated in Figure 14 via a specific example of interest from Figure 4. Figure 14a shows martensite grains with their boundaries in black and the clusters for parent grain reconstruction using Code 8 circled in red. Figure 14b shows that parent grain reconstruction of these clusters results in a parent austenite grain containing an annealing twin. To investigate whether the annealing twin was reconstructed correctly, the fit of the γ variants of all α′ orientation pixels with either, the mean γ orientation and the mean twinned γ orientation was determined. The best fit was used to recalculate the correct γ variant for each pixel and resulted in the orientation map in Figure 14c. The plot reveals that the annealing twin from Figure 14b extends even further and that a second large annealing twin was not detected during parent grain reconstruction at all.

In Figure 14d and e, the $(001)_\gamma$ pole figures of the two γ orientations and their α′ variants according to the K-S and refined ORs are used to reveal the origin of the incomplete reconstruction. The K-S OR dictates that the variants constituting a single martensite packet must satisfy the following conditions:

i. The $\{011\}_{\alpha'}$ planes of all variants in a packet must lie parallel to the same $\{111\}_\gamma$ planes.
ii. The $\langle 111 \rangle_{\alpha'}$ directions on $\{011\}_{\alpha'}$ planes must be parallel to a $\langle 011 \rangle_\gamma$ directions on $\{111\}_\gamma$ planes.

Since the misorientation describing the relationship between γ and twinned γ is defined by a 60° rotation about a $\langle 111 \rangle_\gamma$ axis, the two above conditions can also be satisfied by a single packet of the twinned γ. This packet is hereafter referred to as the "shared packet". The α′ variants of the shared packet are shown by green markers in Figure 14d. Fortunately, as demonstrated in Section 4.1.1, the experimental OR in lath martensitic steels is irrational. The pole figure in Figure 14e shows that in this case, the variants of the shared packet have a misorientation angle of 2.8°. With sufficiently accurate data and a representative refined irrational OR determined from EBSD map data, it should be possible to separate twinned orientations during reconstruction.

The reason for the inaccurate reconstruction of the annealing twins in this example can be found early-on in the procedure. Figure 14a shows that the initial reconstruction of α′ grains with a threshold value of 3° does not separate the variants of the shared packets. Instead, it merges the variants into groups that are known as blocks. It is evident that the unidentified annealing twin in Figure 14d intersects a large block of α′ variants. Therefore, there is no chance of reconstructing this γ annealing twin immediately following the grain reconstruction stage regardless of which grain-level parent grain reconstruction approach is chosen. This is a generic shortcoming of all parent grain reconstruction algorithms. The only available avenues for a more accurate reconstruction of γ annealing twins could be either, a more accurate reconstruction of the α′ variants, or a refinement step on the pixel orientation level as demonstrated in Figure 14c. These avenues will be explored in-depth in future work.

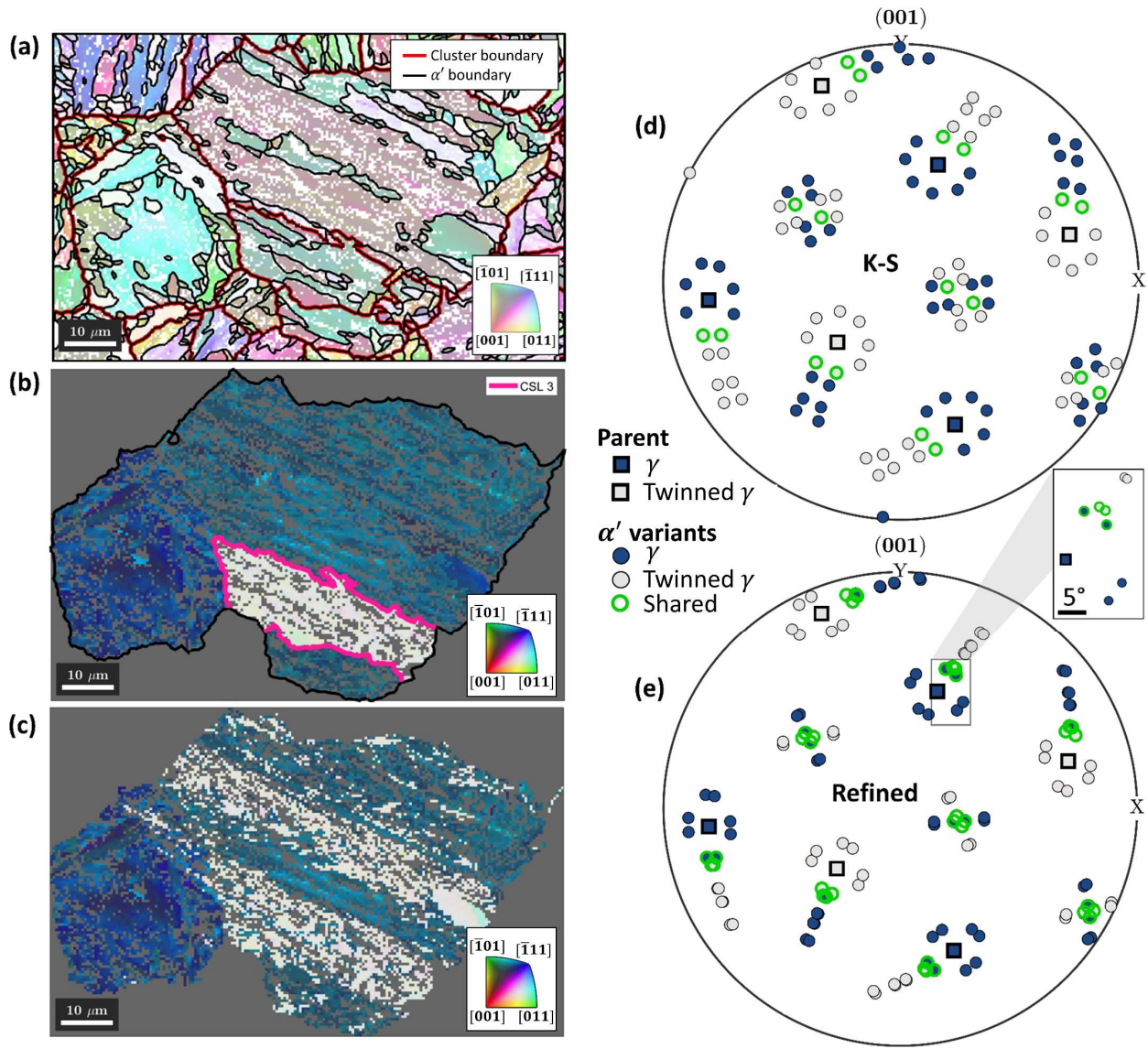

Figure 14. Example of incomplete annealing twin indexing in reconstructed lath martensite (see Section 0). (a) α′ inverse pole figure map, showing grain boundaries in black and clusters formed with Code 8 in red. (b) Reconstructed γ grain containing annealing twin boundaries in pink. (c) Individually best fitting γ orientation for each orientation pixel. (d) $(001)_\gamma$ pole figures of the γ (blue square markers) and twinned γ orientations (grey square markers) from (b). The α′ variants according to the K-S OR are shown by round markers. Variants of packet 1 of γ and packet 4 of twinned γ are shared and shown using green round markers. (e) The variants of the experimentally refined orientation relationship show a 2.8° misorientation between the shared variants.

# 6 Conclusion

This study demonstrates: (i) the implementation of a versatile generic framework, involving a new class, *parentGrainReconstructor*, for parent grain reconstruction from fully or partially transformed child microstructures in the open-source crystallographic toolbox *MTEX* v.5.6 or higher and, (ii) the extension of traditional parent grain reconstruction, phase transformation and variant analysis to all parent-child crystal symmetry combinations.

Three examples of parent grain reconstruction in different transformation microstructures were provided namely, (i) α′-to-γ in a lath martensitic steel, (ii) α-to-β in a Ti alloy, and (iii) a two-step parent grain reconstruction from α′-to-ε-to-γ in a twinning and transformation - induced plasticity steel. The examples showcase the inherent versatility of the universally applicable parent grain reconstruction methods, and the ability to conduct in-depth variant analysis via example workflows that can be programmatically modified by users to suit their specific applications. The latter is significantly simplified by the add-on function library, *ORTools.*

Lastly, for the specific case of austenite annealing twins in martensitic steel, the method to extend the current grain-level parent grain reconstruction approach to pixel orientation level refinement is detailed.

## Acknowledgements

The authors thank Nyyssönen et al. [20], Susanne Hemes, Access e.V. and Pramanik et al. [33] for the EBSD map data of lath martensitic steel, Ti alloy and twinning and transformation - induced plasticity steel.

F. Niessen acknowledges the financial support by the Danish Council for Independent Research grant DFF-8027-00009B. A. A. Gazder acknowledges the 2019 AIIM for Gold - Investigator grant. The JEOL JSM-7001F was funded by the Australian Research Council – Linkage, Infrastructure, Equipment and Facilities grant LE0882613. The Oxford Instruments 80 mm$^2$ X-Max EDS detector was funded via the 2012 UOW Major Equipment Grant scheme.